\documentclass[review,12pt,longtitle]{elsarticle}
\usepackage{graphicx}
\biboptions{sort&compress}
\usepackage{threeparttable}
\usepackage{fancyhdr,afterpage}
\usepackage{amssymb}

\begin{document}

\begin{frontmatter}

\begin{center}

\title{A new acoustic lens material for large area detectors in photoacoustic breast tomography}

\author[rvt]{Wenfeng Xia}
\author[rvt]{Daniele Piras}
\author[rvt]{Johan C. G. van Hespen}
\author[rvt]{Wiendelt Steenbergen}
\author[rvt]{Srirang Manohar\corref{cor1}}\ead{S.Manohar@utwente.nl}

\cortext[cor1]{Corresponding author}

\address[rvt]{Biomedical Photonic Imaging group, Mira Institute
for Biomedical Technology and Technical Medicine, University of
Twente, P.O. Box 217, 7500AE Enschede, The Netherlands.}

\end{center}

\begin{abstract}
\textbf{Objectives:} We introduce a new acoustic lens material for photoacoustic tomography (PAT) to improve lateral resolution while possessing excellent acoustic acoustic impedance matching with tissue to minimize lens induced image artifacts.

\noindent \textbf{Background:} A large surface area detector due to its high sensitivity is preferable to detect weak signals in photoacoustic mammography. The lateral resolution is then limited by the narrow acceptance angle of such detectors.
Acoustic lenses made of acrylic plastic (PMMA) have been used to enlarge the acceptance angle of such detectors and improve lateral resolution. However, such PMMA lenses introduce image artifacts due to internal reflections of ultrasound within the lenses, the result of acoustic impedance mismatch with the coupling medium or tissue.

\noindent \textbf{Methods:} A new lens is proposed based on the 2-component resin Stycast 1090SI. We characterized the acoustic properties of the proposed lens material in comparison with commonly used PMMA, inspecting the speed of sound, acoustic attenuation and density. We fabricated acoustic lenses based on the new material and PMMA, and studied the effect of the acoustic lenses on detector performance comparing finite element (FEM) simulations and measurements of directional sensitivity, pulse-echo response and frequency response. We further investigated the effect of using the acoustic lenses on the image quality of a photoacoustic breast tomography system using k-Wave simulations and experiments.

\noindent \textbf{Results:} Our acoustic characterization shows that Stycast 1090SI has tissue-like acoustic impedance, high speed of sound and low acoustic attenuation. These acoustic properties ensure an excellent acoustic lens material to minimize the acoustic insertion loss. Both acoustic lenses show significant enlargement of detector acceptance angle and lateral resolution improvement from modeling and experiments. However, the image artifacts induced by the presence of an acoustic lens are reduced using the proposed lens compared to PMMA lens, due to the minimization of internal reflections.

\noindent \textbf{Conclusions:} The proposed Stycast 1090SI acoustic lens improves the lateral resolution of photoacoustic tomography systems while not suffering from internal reflection-induced image artifacts compared a lens made of PMMA.

\end{abstract}


\end{frontmatter}

\section{Introduction}

X-ray mammography and ultrasonography, the current routinely used breast imaging methods, have limitations. X-ray mammography besides possessing ionizing hazards, is not reliable for detecting cancer in dense breasts, while ultrasonography suffers from poor soft tissue contrast~\cite{Lehman2007}. Thus there is a need for a breast imaging method, sensitive and specific enough to detect early stages of cancer while using non-ionizing radiation. Optical mammography has the advantage of high optical contrast but suffers from poor spatial resolution caused by multiple light scattering in tissue~\cite{Tromberg2008}. Photoacoustic imaging combines light and sound, where a short laser pulse excites ultrasound by thermo-elastic expansion at locations where light is absorbed. The method has the advantages of both high optical contrast and the high resolution possible with acoustic detection~\cite{Razansky2009, Jose2009, Wang2009, Razansky2011, Zhu2009, Wang2003, Beard2011, Jiang2012}, and shows promise as an alternative imaging method to detect angiogenic markers of breast cancer based on hemoglobin absorption~\cite{Piras2010, Kruger2010, Ermilov2009, Pramanik2008, Heijblom2012, Xi2012, Xie2011,Khokhlova2007,Wang2007,Xing2010}.

In photoacoustic breast imaging there is a requirement to detect extremely weak signals generated upto several centimeter deep in tissue. A large-area detector is preferable due to its higher sensitivity than a small-area detector. On the other hand, a large-area detector has a small angle of acceptance at the designed frequency~\cite{Pierce1989}, while in a tomographic geometry it is required that each US detector's acceptance angle is wide enough to detect signals generated throughout the object for each angular position around the object. The consequence of a narrow acceptance angle is that the lateral resolution of the system is reduced when using a large-area detector~\cite{Xu2003}.  

To use a sensitive large-area detector without compromising its acceptance angle, Li et al~\cite{Li2008} and Pramanik et al~\cite{Pramanik2009} introduced the negative acoustic lens concept. A hemicylindrical-shape negative lens made from acrylic plastic (PMMA) was epoxied to the surface of a large surface-area ultrasound transducer to enlarge the transducer acceptance angle. The lateral resolution of the PAT system was dramatically improved as expected. However, the authors pointed out that image artifacts could result possibly introduced by ultrasound internal reflections within the PMMA lens~\cite{Pramanik2009}.

Materials used as acoustic lenses should provide sufficient refraction to provide lensing effect but with minimum insertion losses. Lensing increases with increase of speed of sound difference between the coupling medium and the lens material, while for low acoustic losses, the material should possess low acoustic reflection and attenuation. Therefore, the ideal acoustic lens material must have a speed of sound $c$ higher than tissue, an acoustic impedance $Z$ as close to that of tissue as possible, and minimal acoustic attenuation (AA). Acrylic plastic (PMMA) due to its high $c$, and low AA remains a choice for acoustic lens material, but acoustic reflection, due to its $Z$ mismatch with soft tissue, causes loss of acoustic signal. Moreover, internal reflections at the boundaries of the lens with tissue causes image degradation~\cite{Brown2007}. 

In this work, we introduce an acoustic lens material with excellent properties as above, based on Stycast 1090SI. We compare this material with PMMA, in terms of $c$, AA and $Z$. The angular sensitivity, pulse-echo and frequency response of a large surface area ultrasound detector with and without acoustic lenses are measured and simulated for the two materials. Finally, we compare the performances of a photoacoustic tomographic system when the two lenses are used affixed on a custom-made single-element detector. Imaging quality improvement is achieved for the Stycast 1090SI lens compared to PMMA lens. We conclude that Stycast 1090SI is an excellent candidate for an acoustic lens used in photoacoustic tomography to enlarge the acceptance angle of a large surface area detector.

\section{Materials and methods}

\subsection{Lens materials}

\subsubsection{Stycast 1090SI}

The resin Stycast 1090SI and the hardener Catalyst 24LV (Henkel, D$\rm\ddot{u}$sseldorf, Germany) are accurately weighed into a clean container in the ratio of 100:23 in weight. The components are blended by hand for 3-5 minutes, and the bottom and side of the mixing container are scraped frequently until no sphere-like clusters can be felt. To remove any entrapped air introduced during mixing, the mixture is degassed under vacuum (1-3 mbar) for about 10-15 minutes. During this operation, the foam rises several times the liquid height. We slightly increase the pressure inside the vacuum to let the foam subside. The foam rising and subsiding procedure is repeated to several times. A silicone based mould release (122S, Emerson \& Cuming, Westerlo, Belgium) is applied to the mould using a soft brush, and the mixture is poured carefully into the mould. Before removing the material out of the mould, it is cast for 24 h at room temperature. 

Two Stycast 1090SI blocks with different thicknesses (dimensions of $15\times15\times2$ mm$\rm ^{3}$ and $15\times15\times4$ mm$\rm ^{3}$) and hemispherical acoustic lenses with diameter of 5 mm were prepared. The diameter of the lens is chosen based on the dimensions of the detector surface area (5 x 5 mm$\rm ^{2}$). The two blocks are used for acoustic characterization and the acoustic lenses are used for studying the transducer angle of acceptance and phantom experiments for photoacoustic tomography.

\subsubsection{Acrylic plastic (PMMA)}

For comparison with Stycast 1090SI lens, PMMA hemispherical lenses are used. These are fabricated by carefully sectioning PMMA balls (4.8 mm diameter, Engineering Laboratories, Inc., Oakland, New Jersey) to get hemispheres . Similarly as for Stycast 1090SI, two PMMA blocks (dimensions of $35\times35\times2.75$ mm$\rm ^{3}$ and $35\times35\times3.50$ mm$\rm ^{3}$) are custom-made to assess the acoustic properties.  

\subsection{The detector}

A custom-made single-element ultrasound detector was fabricated using CTS 3203-HD (CTS Communications Components, Inc., Albuquerque, NM) as the piezoelectric material. The detector with dimensions of active surface 5 x 5 mm$\rm ^{2}$ is optimized to have a high sensitivity (0.5 Pa minimum detectable pressure at 1 MHz) for photoacoustic breast tomography~\cite{Xia2012}. It has a center frequency of 1 MHz with a 80$\%$ fractional bandwidth. The detector is subdiced to reduce the radial resonance caused by the large thickness/width ratio of the active material CTS 3203-HD. With subdicing, the detector is divided into 25 small units with dimensions of 0.9 x 0.9 mm$\rm ^{2}$. The units are grouped electrically but separated ultrasonically by air kerfs. More technical and performance details may be found in Ref.~\cite{Xia2012}.

\subsection{Material acoustic properties characterization methods}

\subsubsection{Speed of sound and acoustic attenuation}

\begin{figure}
  \centering
  \includegraphics[angle=0,width=0.7\linewidth]{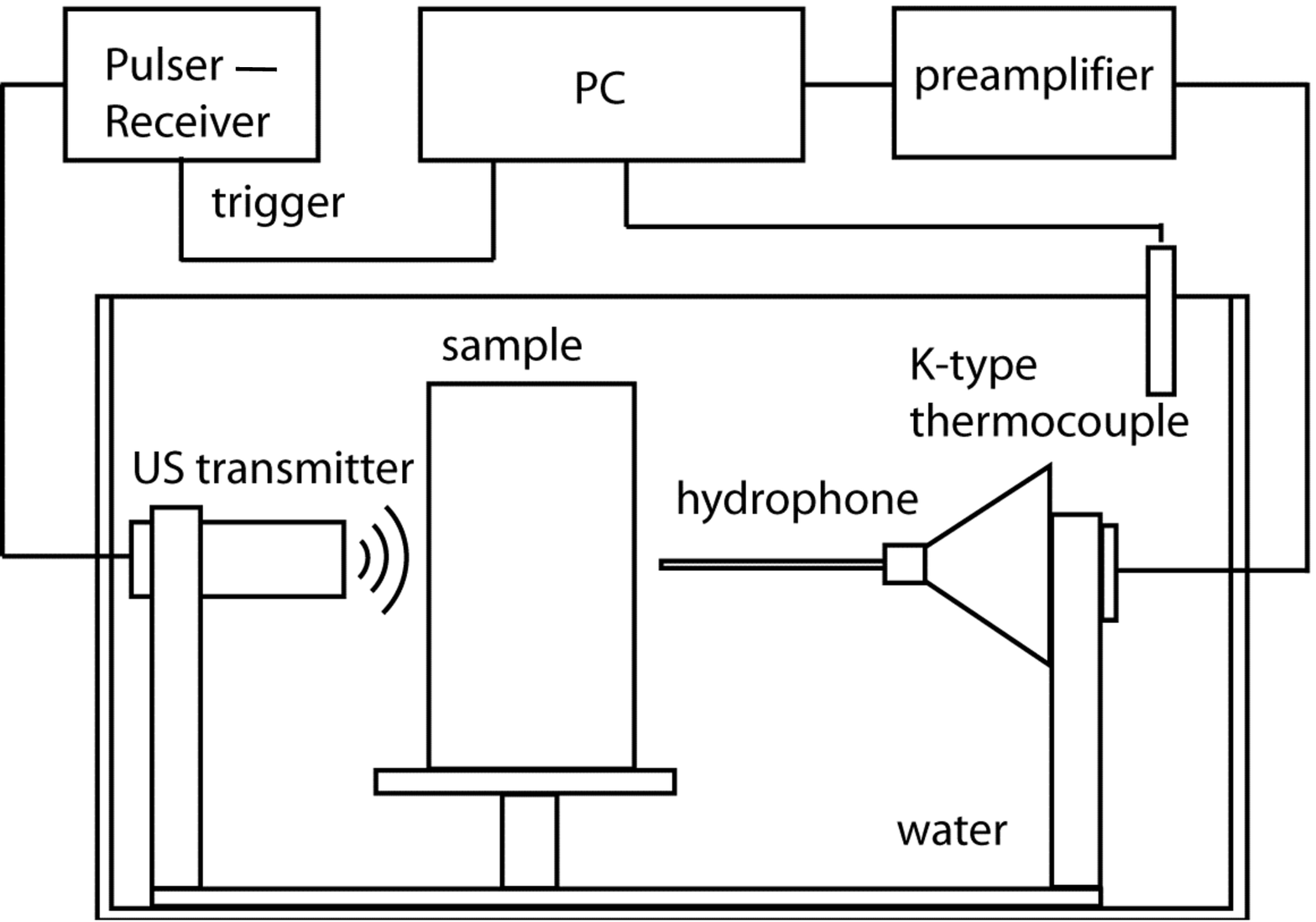}
  \caption{}
  \label{fig:acoustic}
\end{figure}

A modified insertion technique is used to measure the acoustic transmission properties (longitudinal speed of sound and acoustic attenuation)~\cite{Bamber1997}. The system configuration is described in detail elsewhere~\cite{Xia2011}; briefly a single element unfocused Panametrics transducer (for details of central frequencies, see further) together with a broadband needle hydrophone (BLLMCX074 Precision Acoustics Ltd. Dorchester, UK) are mounted in a demineralized water bath at room temperature (Figure~\ref{fig:acoustic}). The transducer is driven by a broadband ultrasonic pulser/receiver (Panametrics 5077PR). The hydrophone measurement data and the water temperature are recorded simultaneously. 

For each material under examination, two samples with different known thicknesses are studied. Signals show different time-of-flights $\Delta T$ in time domain, and amplitude difference $\Delta$A($\omega$) in frequency domain due to the difference in sample thickness $\Delta d$. $\Delta T$ is estimated using cross-correlation of the transmitted signals measured through the two samples. Since the two samples are made of the same material, the acoustic reflections at the water/material interfaces for both samples are equal and canceled during the AA calculation. In this way, inaccuracy in AA estimation caused by acoustic reflection from the sample is minimized. 

Knowing the speed of sound of water $C_{\textrm{w}}$ at the recorded measurement temperature~\cite{Lubbers1998}, the speed of sound $C_{s}$ and the acoustic attenuation $\alpha_{s}$($\omega$) (dB~cm$^{-1}$) for the unknown material can be calculated in time domain and frequency domain respectively by~\cite{Fish1990}: 

\begin{equation}\label{eq1}
\Delta T=\frac{\Delta d}{C_{\mathrm{w}}} - \frac{\Delta d}{C_{s}}
\end{equation}

\begin{equation}\label{eq2}
\alpha_{\mathrm{s}}(\omega) =20 \mathrm{log} \left[ \frac {\Delta  A(\omega)} {\Delta d} \right] + \alpha_{\mathrm{w}}(\omega)
\end{equation}

\noindent where $\alpha_{\mathrm{w}}(\omega)$ the AA of water is negligible since it is small compared to the attenuation of the material~\cite{Greenspan1972}.

For measurement over a wide frequency range, a set of Panametrics transducers with different center frequencies are used: 1, 2.25, 5, 7.5, and 10 MHz covering the range between 0.4 - 12 MHz. AA data are then taken from a frequency range where the calculated FFT of the measured signal has sufficient signal to noise ratio (SNR) for each transducer. The $c$ and AA are measured 5 times for the proposed acoustic lens material Stycast 1090SI and for PMMA. 

\subsubsection{Density and acoustic impedance}

The density $\rho$ of the proposed lens and of acrylic plastic (PMMA) are measured using a standard pycnometer with deionized water at room temperature as reference. The accuracy of the analytical balance (Sartorius BP 210D, Goettingen, Germany) used for measuring the weight is 0.0001 g. Five measurements are made for each material. The acoustic impedance $Z$ is then calculated by~\cite{Fish1990}:

\begin{equation}\label{eq3}
Z = \rho C_{\textrm{s}}
\end{equation}

\noindent where $C_{s}$ is the speed of sound of the material whose values are taken from the measurement results in section 2.3.1.

\subsection{Detector performance characterization methods}

To study the effect of using acoustic lenses on detector performance, a thin layer of ultrasound gel is sandwiched between the lens and detector. A layer of agar solution is then poured on top of the lens to be gelled for the lens stabilization.

\subsubsection{Directivity (simulation and experiment)}

The acceptance angle of the transducer is simulated and measured in transmission mode, assuming the validity of the reciprocity principle for the piezoelectric material~\cite{Callerama1979}. Simulations are performed using 3D finite-element (FEM) based models using the PZFlex software (version 3.0, Weidlinger Associates Inc, Los Altos, CA). The detailed description of the detector model and material properties of the detector used in simulations are presented in Ref.~\cite{Xia2012}. The properties of the lens materials used in the simulation are listed in Table~\ref{table:acoustic}. 

\begin{figure}
  \centering
  \includegraphics[angle=0,width=1\linewidth]{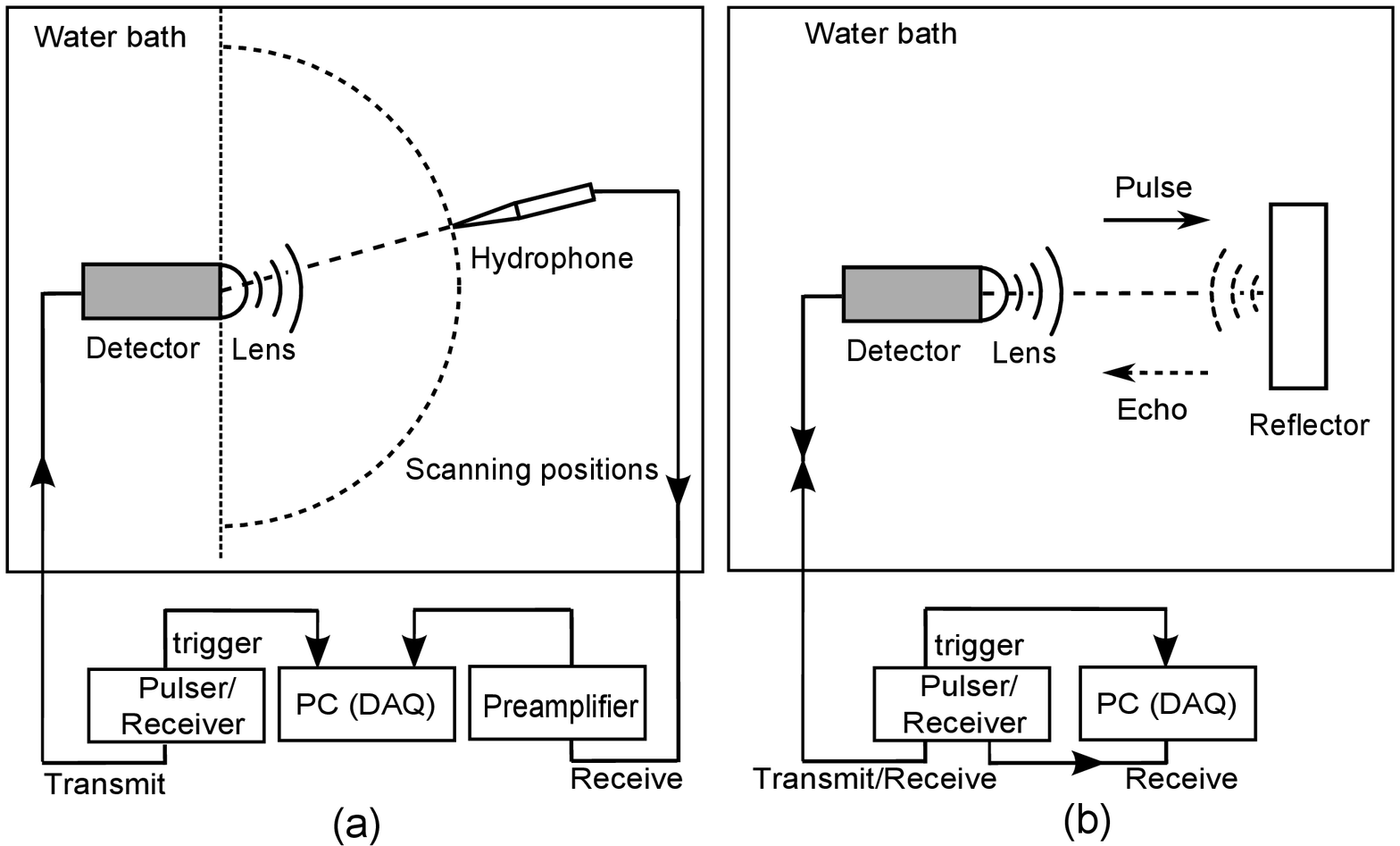}
  \caption{}
  \label{fig:Directivity_setup}
\end{figure}

For the experimental characterization of the acceptance angle, the detector is mounted in a demineralized water bath at room temperature as transmitter driven by a broadband pulser/receiver (Panametrics 5077PR). A broadband needle hydrophone (BLLMCX074 Precision Acoustic Ltd. Dorchester, UK) scans the acoustic field generated by the transducer in the plane covering 180$^{\rm{o}}$ in steps of 1$^{\rm{o}}$ as shown in Figure~\ref{fig:Directivity_setup}(a). The acoustic signals are recorded with a data acquisition card (U1067A Acqiris, 8 bit, 500 MS/s). Detector acceptance angles without lens, with Stycast 1090SI lens and PMMA lens are measured.

\subsubsection{Frequency response (simulation and experiment)}

The frequency response of the detector, with and without acoustic lenses, is simulated and measured in pulse-echo mode (Figure~\ref{fig:Directivity_setup}(b)). Simulations are performed using 3D FEM models.

For experiments, the detector is driven by a broadband ultrasonic pulser/receiver (Panametrics 5077PR). A stainless steel plate is placed in the far-field ($\sim$55 mm) of the detector as an acoustic reflector. The frequency response of the detector is calculated by the square root of the FFT of the measured pulse-echo signal to account for the two-way response.

\subsection{Acoustic lenses used in photoacoustic tomography experiments}

\subsubsection{Forward problem (simulation and experiment)}

\begin{figure}
  \centering
  \includegraphics[angle=0,width=1\linewidth]{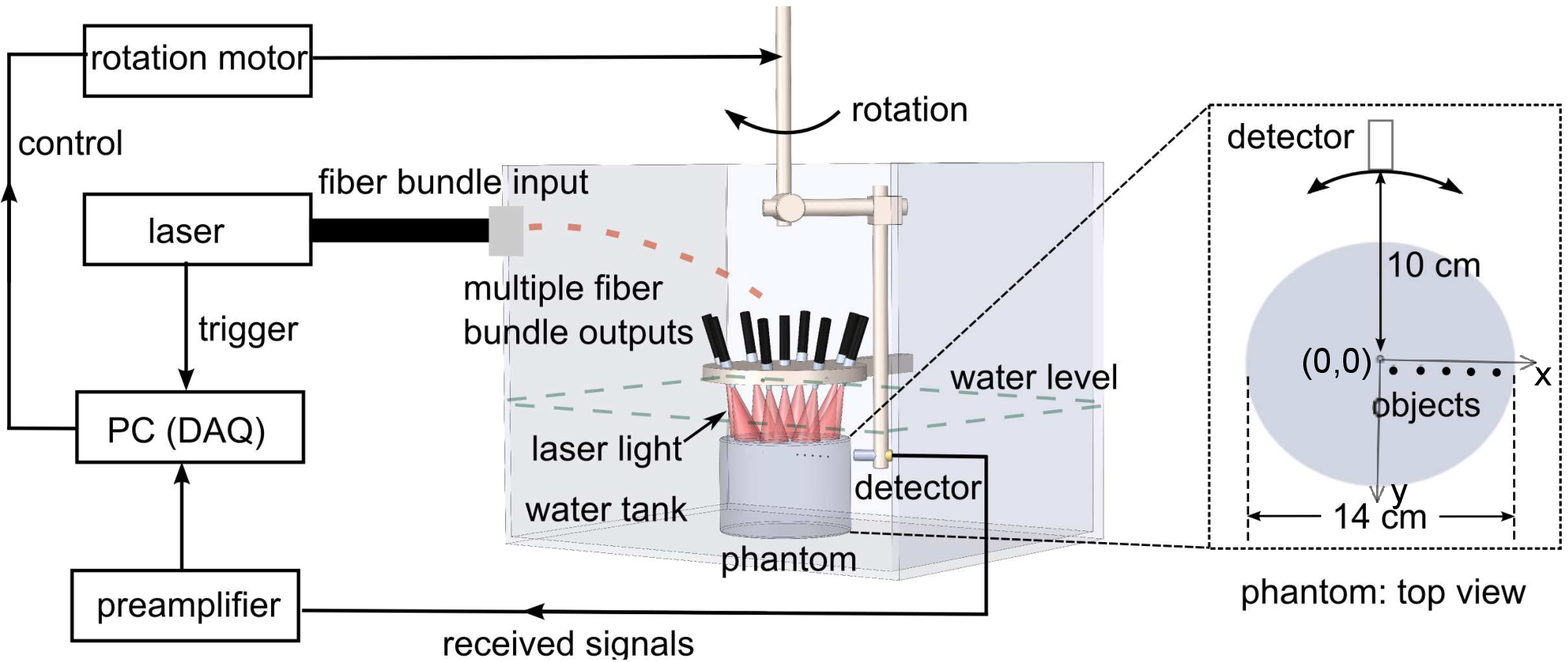}
  \caption{}
  \label{fig:CT_setup}
\end{figure}

The same phantom and tomographic design are used for both simulation and experiment. The phantom has a cylindrical shape (14 cm in diameter, with optical absorption $\mu_{\textrm{a}}$ = 0.003~$\rm{mm^{-1}}$, reduced optical scattering $\mu_{\textrm{s}}^{'}$ = 0.4~$\rm{mm^{-1}}$). Five highly absorbing spherical objects ($\mu_{\textrm{a}}$ = 33~$\rm{mm^{-1}}$, $\mu_{\textrm{s}}^{'}$ = 0.4~$\rm{mm^{-1}}$) with diameters of 2 mm are embedded 1 cm deep in the phantom for 2D imaging as shown in Figure~\ref{fig:CT_setup}. The detector scans the phantom over 240$\rm{^{o}}$ with scanning step of 2$\rm^{o}$ with radius of 10 cm. A complete circular scan could not be performed due to mechanical constraints in the form of a holder for the detector. Three experiments are performed using (1) the bare detector, (2) detector with PMMA lens and (3) detector with Stycast 1090SI lens. 

The simulations are performed using the k-Wave toolbox~\cite{Cox2005,Treeby2010s}. A 2D initial pressure distribution map (1024 x 1024 grid, 20 x 20 cm$\rm ^{2}$ size) is assigned in a 2D tomographic configuration. A initial pressure value of 1 is assigned to the objects, a pressure value of zero is given to background. Measured $c$, $\rho$ and AA of the lens material and medium (water) are assigned (Table~\ref{table:acoustic}) in the model. A 5 mm detector is attached to the lens for acoustic wave recording. To simulate the acoustic lens induced internal reflection, $c$ and $\rho$ of the detector front matching layer (FML) are assigned to a 5 x 10 mm$\rm ^{2}$ area with appropriate AA directly behind the k-Wave virtual detector. The time-domain PA signals are averaged over the surface of the detector and then convolved with the measured impulse response of the detector, which is presented in Ref.~\cite{Xia2012}. 

For experiments, laser pulses with 10 Hz repetition rate at 755 nm wavelength (QCombo, Quanta System, Milan, Italy) are delivered via a fiber bundle (1 input : 9 outputs) to illuminate the phantom from the top (Figure~\ref{fig:CT_setup}). The input diameter is 9 mm, and each of the output has a diameter of 3.5 mm (LGO Optics, San Jose, CA). The phantom and objects are made of agar/Intralipid (2$\%$ agar in weight, 0.3$\%$ Intralipid in volume~\cite{Staveren1991, Curcio1951, Hale1973, Xia2011, Martelli2007}), with the object having additonal 5$\%$ India ink for contrast~\cite{Martelli2007}). The phantom is placed in demineralized water in an imaging tank. The laser beam from each fiber output ($\sim$9 mJ) is naturally diverged to cover the entire phantom surface, resulting in less than 1 mJ~cm$^{-2}$ fluence at phantom surface. The detector is situated at a distance of 10 cm from the center of the phantom and rotated to aquire projections mentioned earier. The signals are amplified using a custom-made low noise preamplifier based on ADA4896-2 (Analog Devices, Norwood, MA). The signals are then digitized and transferred (U1067A Acqiris, 8 bit, 500 MS/s) to a PC for data processing. 

\subsubsection{Image reconstruction}

For both simulations and experiments, k-Wave toolbox is employed for image reconstruction using time-reversal~\cite{Treeby2010s, Treeby2010}. A different grid (1000 x 1000) is used for reconstruction compared to the grid used in the forward simulations (1024 x 1024) to avoid the inverse crime. The $c$ of the lens is taken into account by assigning the value to the lens area. To avoid the generation of internal reflections in the lens during the time-reversal reconstruction, the density of each lens is tuned to match the impedance of water and the 5 x 10 mm$\rm ^{2}$ high impedance region used in the forward simulation is removed. 

\subsubsection{Image contrast analysis} 

To estimate the image contrast, for each reconstructed object, a 1 x 1 cm$\rm ^{2}$ image region of interest (ROI) is considered with object located in its center. The object in the image is defined by including pixels with values upwards of -6 dB of the peak intensity, while the area outside this region is defined to be noise. A contrast to noise ratio (CNR) for each object is calculated as the ratio of the root-mean-square (RMS) value of the object and noise pixels. 

\section{Results}

\subsection{Material acoustic properties}

\begin{figure}
  \centering
  \includegraphics[angle=0,width=1\linewidth]{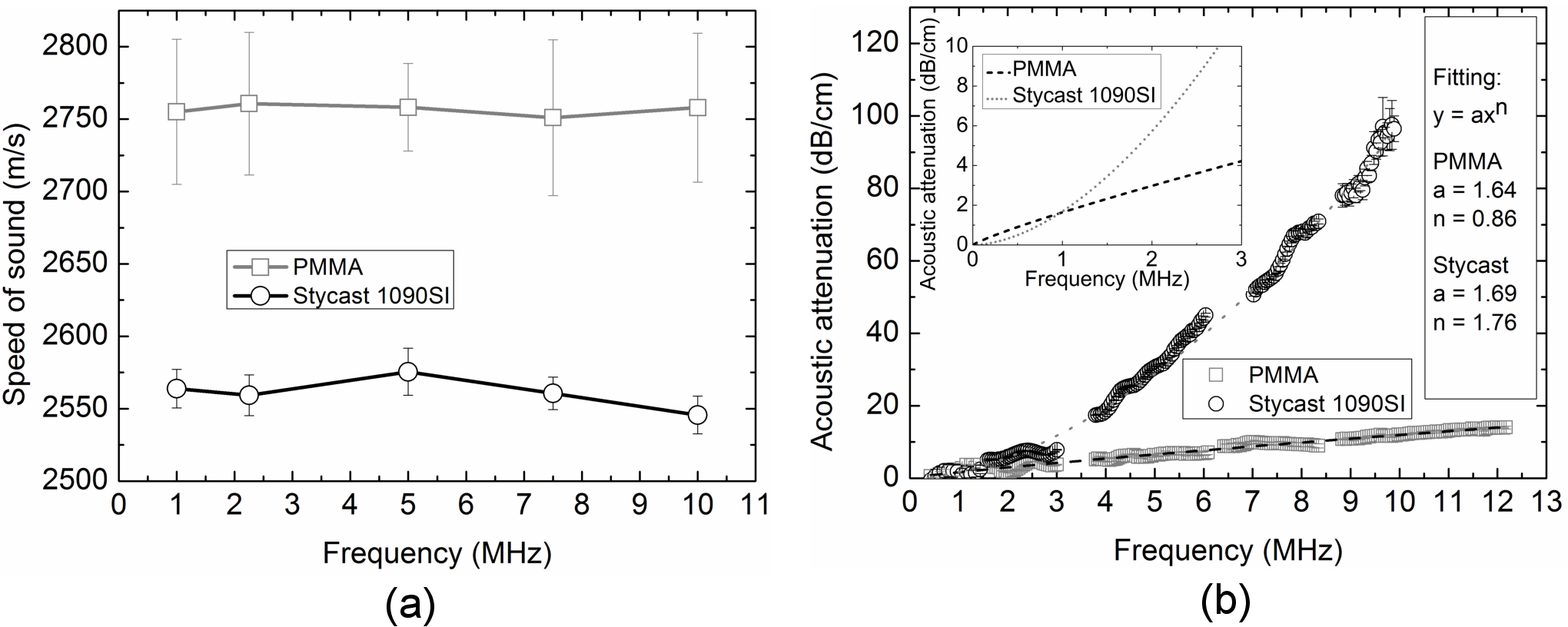}
  \caption{}
  \label{fig:acousticresults}
\end{figure}

Figure~\ref{fig:acousticresults} shows the measured acoustic transmission properties at different ultrasound frequencies for Stycast 1090SI and acrylic plastic (PMMA) blocks. In general, our measurement results for PMMA agree well with literature values ~\cite{Bloomfield2000,Carlson2003}, indicating the accuracy of our acoustic characterization. The measured $c$ (Figure~\ref{fig:acousticresults}(a)) is slightly higher for PMMA than for Stycast 1090SI, while both have much larger $c$ than water at the measurement temperature of 21$^{0}$C, indicating an efficient enlargement of the transducer acceptance angle when used as acoustic lenses according to Snell's law. For both materials, the $c$ is constant for different frequencies, and no dispersion trend can be observed though relatively large standard deviations are observed. Larger standard deviations of the measured $c$ values for PMMA are registered compared to those for Stycast 1090SI. This is because the $\Delta d$ for the two PMMA blocks (0.75 mm) is smaller than that for Stycast 1090SI (2 mm); the uncertainty in thickness estimation has larger influence on $c$ estimation for PMMA. 

The measured AA (Figure~\ref{fig:acousticresults}(b)) has low standard deviations and follows the frequency power law as for most materials~\cite{Waters2005}. The AA for Stycast 1090SI and PMMA can then be expressed from a power law fit to frequency (f, in MHz) as 1.64$\rm f^{0.86}$ and 1.69$\rm f^{1.76}$ respectively. However for Stycast 1090SI, higher variations are presented at high frequencies due to the large attenuation of the material. Both Stycast 1090SI and PMMA have low AA at low frequency regime (around 1 MHz), while its value increases with frequency more rapidly for Stycast 1090SI than for PMMA. 

The standard deviations for $c$ values are relatively large because thin samples are used where there is uncertainty in thickness estimation. The reason for using thin samples is to avoid the heavy attenuation and thus to have high SNR for the frequency domain analysis of the detected signals. 

The measured $\rho$ of Stycast 1090SI (0.56 gcm$^{-3}$) is lower than that of water, and combined with the high $c$, Stycast 1090SI has $Z$ (1.43 MRayl) close to tissue (water), while PMMA has higher $Z$ (3.32 MRayl) than water (Table~\ref{table:acoustic}).

\subsection{Acceptance angle using the acoustic lenses}

\begin{figure}
  \centering
  \includegraphics[angle=0,width=1\linewidth]{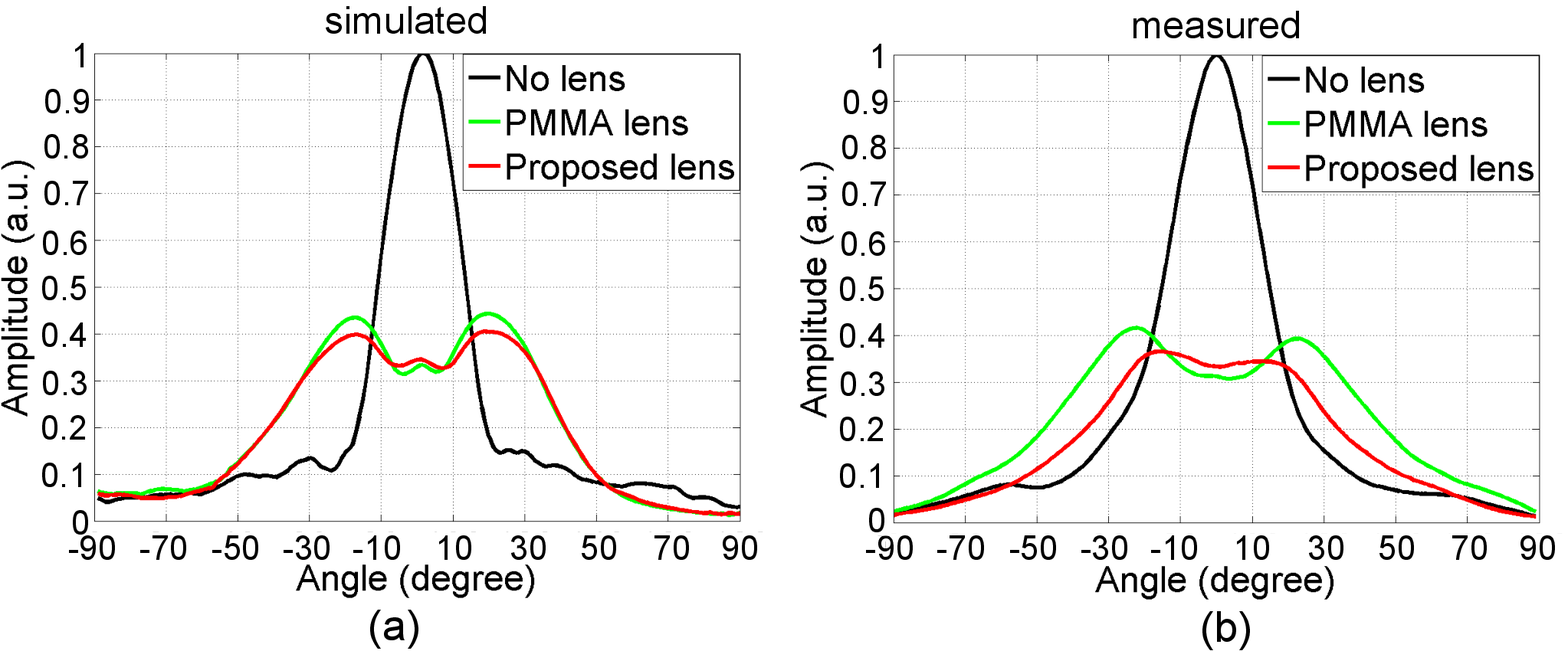}
  \caption{}
  \label{fig:directivity_results}
\end{figure}

The simulated and measured directional sensitivity of the detector with and without lenses are shown in Figure~\ref{fig:directivity_results}. In general, the simulation results match the experimental results. The -6 dB acceptance angle of the bare detector is 25$\rm^{o}$ (20$\rm^{o}$ from simulation). Both simulated and measured directional sensitivity curves for the detector with lenses have a valley in the center region. The contribution from the uncovered areas enhanced the sensitivity of the detector in front of those areas, forming the valleys in the curves. Since the $c$ of PMMA is larger than Stycast, the acceptance angle of the detector with PMMA lens (95$\rm^{o}$ from measurement, 85$\rm^{o}$ from simulation) is larger than Stycast(75$\rm^{o}$ from measurement, 80$\rm^{o}$ from simulation). 

\subsection{Pulse-echo and frequency response}

\begin{figure}
  \centering
  \includegraphics[angle=0,width=1\linewidth]{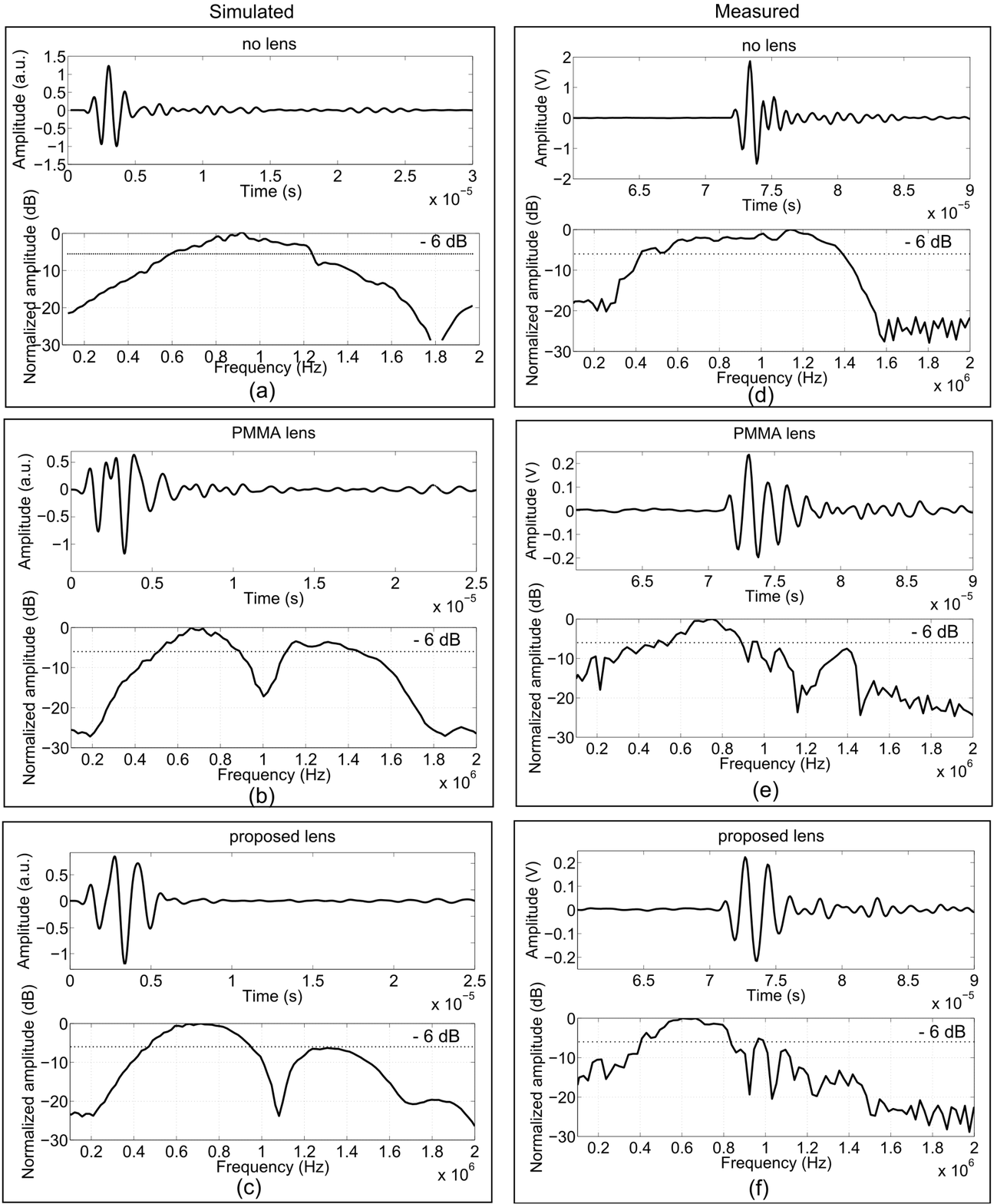}
  \caption{}
  \label{fig:pulse_echo}
\end{figure}

Pulse-echo and frequency response of the detector with and without lenses are simulated and compared with measurement results in Figure~\ref{fig:pulse_echo}. In general, simulations agree well with measurements. The bare detector has a short pulse length with 1 MHz center frequency and $\sim$80$\%$ fractional bandwidth (Figure~\ref{fig:pulse_echo}(a)and(d)). When the PMMA lens is attached to the detector, the pulse length increases due to the internal reflection inside the PMMA lens; the bandwidth of the detector falls to 50$\%$ (Figure~\ref{fig:pulse_echo}(e)). The center frequency is also reduced to be 700 kHz (Figure~\ref{fig:pulse_echo}(b)and(e)). As for the case of Stycast 1090SI lens, the pulse-echo has the same number of cycles as for the bare detector (3 cycles), and shorter than PMMA lens (more than 4 cycles). The center frequency shifts to 600 kHz, while the bandwidth is only slightly reduced to be $\sim$75$\%$ compared to the bare detector (Figure~\ref{fig:pulse_echo}(f)). 

\subsection{Imaging quality}

\begin{figure}
  \centering
  \includegraphics[angle=0,width=0.7\linewidth]{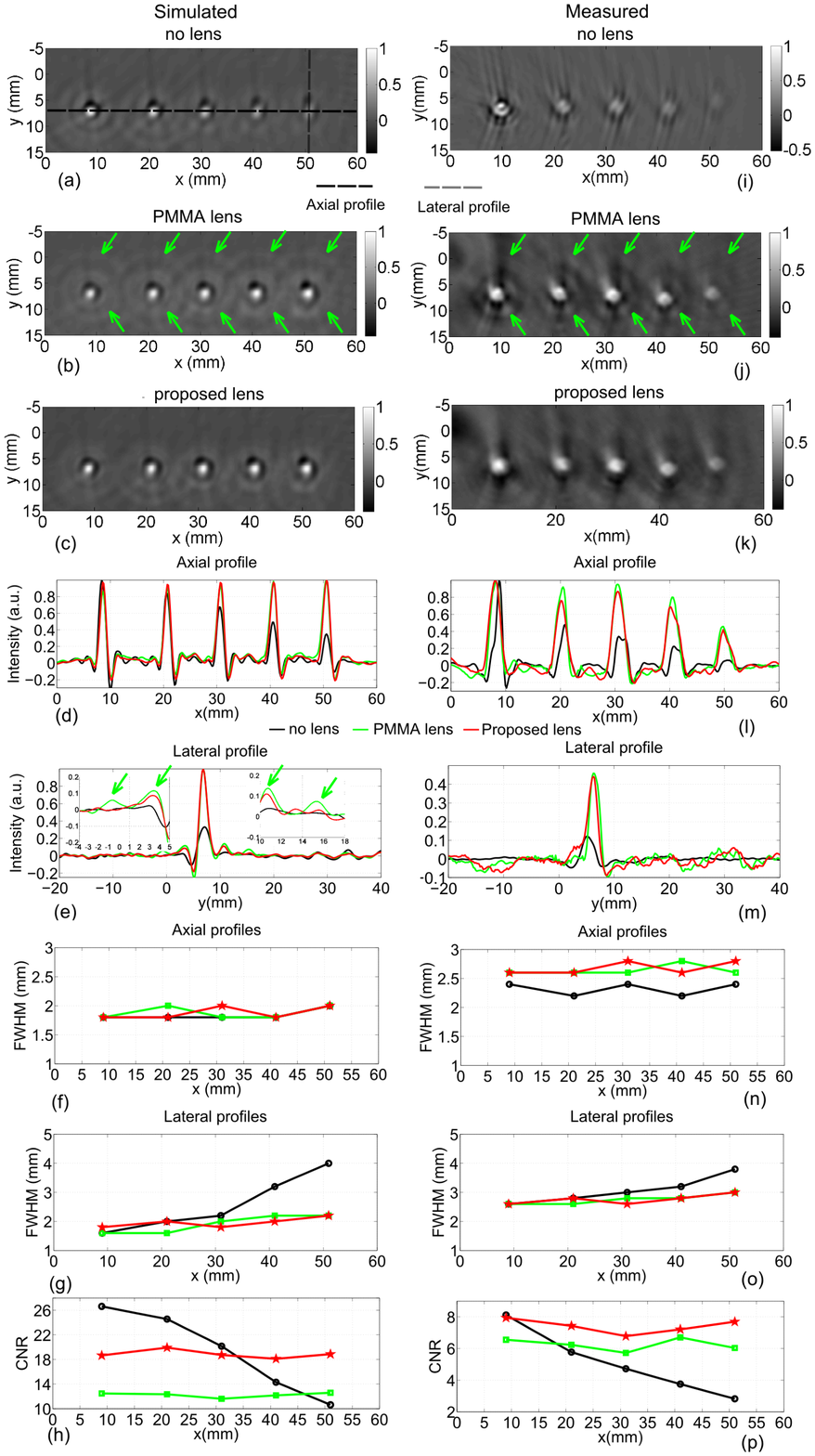}
  \caption{}
  \label{fig:CTmeasurement}
\end{figure}

Figure~\ref{fig:CTmeasurement} shows the reconstructed images from PAT simulations (left column) and measurements (right column). In general, simulations agree with the measurements. Reconstructions of the objects imaged with the bare detector are shown in Figure~\ref{fig:CTmeasurement}(a)and(i). The objects are less and less visible and elongated in the Y direction when moving away from the center of rotation at x=0, y=0. On the other hand, the reconstructions of the objects imaged with both acoustic lenses possess better image quality at all x: all the objects are clearly visible with well preserved shapes (Figure~\ref{fig:CTmeasurement}(b),(c)and(j),(k)). The axial and lateral profiles for all objects are shown in Figure~\ref{fig:CTmeasurement}(d),(i), and (e), (m) respectively. The full width at half maximum (FWHM) values of the axial and lateral profiles for each object imaged with and without lenses are shown in Figure~\ref{fig:CTmeasurement}(f)(g)(n) and (o). They indicate that the axial resolution remains the same when the objects are imaged with acoustic lenses attached to the detector as compared to the bare detector, while the lateral resolution improved, with a FWHM reduced from $\sim$4 mm to $\sim$2 mm using both PMMA and Stycast 1090SI lenses as expected. We also calculated contrast to noise ratio (CNR) for each object to appreciate the differences in performance between the Stycast 1090SI and PMMA lenses. The CNR was calculated as the ratio of the root-mean-square (RMS) value of the object and noise pixels, with the boundary between the two regions the -6 dB points of highest pixel intensity in a ROI around each object. The CNRs of each reconstructed object are shown in Figure~\ref{fig:CTmeasurement}(h) and (p).  

Objects imaged with the bare detector have ring-shape artifacts due to the limited bandwidth of the detector (Figure~\ref{fig:CTmeasurement}(a)and(i)). Those ring-shape artifacts are even enhanced for objects imaged by the detector with PMMA lens due to the reverberation of ultrasound inside the lens (indicated by green arrows in Figure~\ref{fig:CTmeasurement}(b)and(j)). This is because the acoustic impedance of PMMA lens lies well between the front matching layer and surrounding water/tissue. On the other hand, objects imaged by Stycast 1090SI lens show much weaker artifacts as compared to PMMA lens. The reduction of image artifacts can be clearly seen from the lateral profiles for the object close to the detection side in Figure~\ref{fig:CTmeasurement}(e)insets. Further the CNRs of objects imaged by the Stycast 1090SI lens are higher than those from the PMMA lens, due to the reflection artifacts associated with the latter (Figure~\ref{fig:CTmeasurement}(h) and (p)).

\section{Discussion}
Li et al~\cite{Li2008} introduced the concept of using an acoustic lens to enlarge the narrow acceptance angle of a large surface-area detector, and consequently improve the lateral resolution of the system. Later Pramanik et al~\cite{Pramanik2009} applied this concept in their photoacoustic/thermocoustic breast imager using an acoustic lens made of PMMA. Due to the acoustic impedance mismatch between PMMA and detector front matching layer, and between PMMA and the surrounding medium, reverberation of ultrasound inside the lens could cause image artifacts as pointed out by the authors~\cite{Pramanik2009}.

We introduced a new acoustic lens material Stycast 1090SI, which has a remarkable tissue-like acoustic impedance, which is more suitable than PMMA for use as an acoustic lens in photoacoustic tomography. We detailed the fabrication of the material. Care needs to be taken to prevent air bubbles entrapped inside the material during material manufacturing. We studied the effect of using this acoustic lens on the detector performances and on the image quality of the system. We find that besides improved lateral resolution compared to bare detector, image artifacts are minimized by using the proposed lens compared with PMMA. This is because the material has large speed of sound to enlarge the detector acceptance angle, while having tissue-like acoustic impedance to minimize the ultrasound internal reflections within the lens.  

The acoustic attenuation of Stycast 1090SI increases rapidly with frequency (Figure~\ref{fig:acousticresults}(b). For frequencies below 1 MHz the acoustic attenuation of the 5 mm diameter (2.5 mm thickness) lens is acceptable making the lens well suited.  For frequencies larger than 3 MHz, the acoustic attenuation within 5 mm diameter lens is $\sim$6 dB. Thus , for higher frequency ultrasound detectors it is may be preferable to use PMMA lenses which possess lower insertion losses, which however induce more image artifacts compared to Stycast 1090SI lens. 

The center frequency of the detector is lowered due to the acoustic attenuation of both PMMA lens and Stycast lens (Figure~\ref{fig:pulse_echo}). It is important to take this into account when designing the center frequency of ultrasound detectors. The bandwidth of the detector is reduced by using PMMA due to the internal reflections within the lens, while less bandwidth reduction is found for Stycast lens compared to PMMA lens.

The directivity curves of the detector with both Stycast 1090SI and PMMA lens attached have a valley in the center (Figure~\ref{fig:directivity_results}). This is because the detector surface area (5 x 5 mm$\rm ^{2}$) is not completely covered by the lens (5 mm diameter). A larger lens (7 mm diameter) covering the entire surface area of the detector is found to solve this problem in simulations (not shown in this work). However, using a larger acoustic lens increases the complexity for the detector array development as part of the lens requires to be removed. Since good image quality has been achieved as shown in Figure~\ref{fig:CTmeasurement}, we decided to use an acoustic lens with 5 mm diameter.   

Interestingly the insertion loss for PMMA lens is smaller than for the Stycast 1090SI lens. This is because the acoustic impedance of PMMA lies between water and the front matching layer of the detector, which makes PMMA function as a second impedance matching layer between the detector front matching layer and surrounding water, increasing the detector sensitivity.  For Stycast 1090SI lens, due to the large difference between the acoustic impedance of the detector front matching layer and the lens material, the reflection loss is large. To prevent the large reflection loss for Stycast 1090SI lens, the acoustic impedance of the detector front matching layer requires to be tuned lower than the present value. In future, detectors with two or more front matching layer will be designed with acoustic impedance of the outer matching layer close to tissue~\cite{McKeighen1998}. In this way, acoustic reflection loss induced by Stycast 1090SI lens will be minimized which will further increase the detector bandwidth.

\section{Conclusions}
Stycast 1090SI due to its tissue-like acoustic impedance, low acoustic attenuation and high speed of sound, is an excellent acoustic lens material for photoacoustic imaging at around 1 MHz detection frequencies. Combined with the large surface area ultrasound detector, acoustic lenses made from Stycast 1090SI can enlarge the acceptance angle of the detector, and thus improve the lateral resolution of the photoacoustic tomographic system for breast imaging with minimized image artifacts, compared with earlier used PMMA.

\section*{Conflict of Interest statement}

The authors declare that there are no conflicts of interest.

\section*{Acknowledgments}

The financial support of the Angentschap NL Innovation--Oriented Research Programmes Photonic Devices under the HYMPACT Project(IPD083374) and High Tech Health Farm, MIRA institute for Biomedical Technology and Technical Medicine are gratefully acknowledged. Spiridon~van~Veldhoven and Christian~Prins from Oldelft Ultrasound B.V, The Netherlands are acknowledged for their contribution in developing acoustic lenses.

\newpage

\section*{List of Figure Captions}

{\sffamily \textbf{Figure 1:} Schematic of the setup for the acoustic transmission properties measurements.}

{\sffamily \textbf{Figure 2:} Schematics of (a) directivity measurement setup and (b) pulse-echo signal and electrical-acoustical frequency response measurement setup.}

{\sffamily \textbf{Figure 3:} Schematic of the measurement configuration for the lenses used in a photoacoustic tomographic system.}

{\sffamily \textbf{Figure 4:} Measured acoustic transmission properties. (a) Speed of sound and (b) acoustic attenuation in Stycast 1090SI and acrylic plastic (PMMA) with frequency power law fitting as insertion for attenuation in low frequency regime. The values are measured at 21 $^{0}$C.  Each data point represents an average of 5 measurements and error bars represent standard deviations of the measured values.}

{\sffamily \textbf{Figure 5:} Directional sensitivity of the detector without a lens, with PMMA lens and with proposed lens: (a) simulated and (b) measured.}

{\sffamily \textbf{Figure 6:} Simulated (left column) and measured (right column) pulse-echo and frequency response of the detector: (a) without lens, (b) with PMMA lens, (c) with Stycast 1090SI lens, (d) without lens, (e) with PMMA lens and (f) with Stycast 1090SI lens. (a)-(c) are simulation results signals time-shifted to origin and (d)-(f) are measurement results.}

{\sffamily \textbf{Figure 7:} Simulated (left column) and experimental (right column) PAT results for the described phantom: (a,i) Objects imaged by detector without an acoustic lens attached, (b,j) by detector with PMMA lens and (c, k) with Stycast 1090SI lens. Each image is normalized to its maximum intensity. (d, l) Axial intensity profiles along the trajectory indicated by dashed black line from (a)-(c). (e, m) Lateral intensity profiles indicated by dashed gray line from (i)-(k). (a)-(h) are simulation results, (i)-(p) are measurement results.  Lateral resolution is improved by using both PMMA and Stycast lenses as expected, while axial resolution remains constant for all three cases. Due to limited bandwidth of the detector, ring-shape artifacts are presented around the objects. Those ring-shape artifacts are enhanced for objects imaged by detector with PMMA lens due to the ultrasound trapped inside the lens (indicated by arrows), while for the objects imaged by the detector with Stycast 1090SI lens, much weaker such artifacts are presented due to the tissue-like acoustic impedance of the lens material.}

\newpage
\section*{List of Table Captions}

{\sffamily \textbf{Table 1:} Measured acoustic properties of lens materials (Stycast 1090SI and PMMA) compared with literature values for water. All values are for material at 21 $^{0}$C and for frequency of 1 MHz. The longitudinal velocity, shear velocity and acoustic attenuation are used in 3D FEM simulations.}

\newpage
\vspace{80mm}
\begin{center}
\end{center}
\newpage

\vspace{80mm}
\begin{center}
\end{center}
\newpage

\vspace{80mm}
\begin{center}
\end{center}
\newpage

\vspace{80mm}
\begin{center}
\end{center}
\newpage

\begin{center}
\end{center}

\begin{center}
\end{center}

\begin{center}
\end{center}
\newpage

\begin{table}
\caption{\label{table:acoustic}}\centering 
\begin{threeparttable}[b]
\begin{tabular}{| c | c | c | c | c | }
\hline
Acoustic properties & PMMA & Stycast 1090SI & Water & FML$^{\rm a}$\\
\hline
Longitudinal velocity (m~s$^{-1}$)&  2757 & 2561 & 1486$^{\rm b}$ & 2068 \\
\hline
Shear velocity (m~s$^{-1}$)&  1400$^{\rm c}$ & 900 & - & 994 \\
\hline
AA (dB~cm$^{-1}$) & 1.64  & 1.69 & - & - \\
\hline
Density (g~cm$^{-3}$) & 1.20  & 0.56  & 1.00$^{\rm }$ & 3.14\\
\hline
Impedance (MRayl) & 3.32 & 1.43 & 1.49$^{\rm b}$ & 6.49\\
\hline
\end{tabular}
\begin{tablenotes}
\item[a] FML: Detector front matching layer, values from Ref.~\cite{Xia2012}.   
\item[b] Ref.~\cite{Lubbers1998}.  
\item[c] Ref.~\cite{Sutherland1978}  

\end{tablenotes} 
\end{threeparttable}
\end{table}
\newpage


\begin{thebibliography}{00}

\bibitem{Lehman2007}
C.~D.~Lehman, C.~Isaacs, M.~D.~Schnall, E.~D.~Pisano, S.~M.~Ascher, P.~T.~Weatherall, D.~A.~Bluemke, D.~J.~Bowen, P.~K.~Marcom, D.~K.~Armstrong, S.~M.~Domchek, G.~Tomlinson, S.~J.~Skates and C.~Gatsonis, ``Cancer yield of mammography, MR, and US in high-risk women: Prospective multi-institution breast cancer screening study,'' {\em Radiology} ~{\bf 244}, (2007), 381-388

\bibitem{Tromberg2008}
B.~Tromberg, B.~W.~Pogue, K.~D.~Paulsen, A.~G.~Yodh, D.~A.~Boas and A.~E.~Cerussi, ``Assessing the future of diffuse optical imaging technologies for breast cancer management,'' {\em Med. Phys.}~{\bf 35} (2008), 2443-2451

\bibitem{Razansky2009}
D.~Razansky, M.~Distel, C.~Vinegoni, R.~Ma, N.~Perrimon, R.~W.~K$\rm\ddot{o}$ster and V.~Ntziachristos, ``Multispectral opto-acoustic tomography of deep-seated fluorescent protein it in vivo,'' {\em Nat. Photonics}~{\bf 3}, (2009), 412-417 

\bibitem{Jose2009}
J.~Jose, S.~Manohar, R.~G.~M.~Kolkman, W.~Steenbergen and T.~G.~van Leeuwen, ``Imaging of tumor vasculature using Twente photoacoustic systems,'' {\em J. Biophoton} ~{\bf 2} (2009), 701-717

\bibitem{Wang2009}
L.~V.~Wang, ``Multiscale photoacoustic microscopy and computed tomography,'' {\em Nat. Photonics}~{\bf 3} (2009), 503-509

\bibitem{Razansky2011}
D.~Razansky, A.~Buehler, and V.~Ntziachristos, ``Volumetric real-time multispectral optoacoustic tomography of biomarkers,'' {\em Nat. Protocols}~{\bf 6} (2011), 1121-1129. doi:10.1038/nprot.2011.351

\bibitem{Zhu2009}
J. Gamelin, A. Maurudis, A. Aguirre, F. Huang, P. Guo, L. V. Wang, and Q. Zhu, ``A real-time photoacoustic tomography system for small animals,'' {\em Opt. Express}, {\bf 17}(13), (2009), 10489-10498 

\bibitem{Wang2003}
X. Wang, Y. Pang, G. Ku,  X. Xie,  G. Stoica, and L. V. Wang, ``Noninvasive laser-induced photoacoustic tomography for structural and functional {\it in vivo} imaging of the brain,'' {\em Nat. Photonics} {\bf 21}, 7, (2003), 803-806

\bibitem{Beard2011}
P.~Beard, ``Biomedical photoacoustic imaging,'' {\em Interface Focus}~{\bf 1}, (2011), 602-631

\bibitem{Jiang2012}
L. Xiang, B.~Wang, L.~Ji, and H.~Jiang, ``4-D photoacoustic tomography,'' {\em Scientific reports}~{\bf 3}, (2012), 1113 doi:10.1038/srep01113

\bibitem{Piras2010}
D.~Piras, W.~Xia, W.~Steenbergen, T.~G.~van Leeuwen and S.~Manohar, ``Photoacoustic imaging of the breast using the Twente Photoacoustic Mammoscope: Present status and future perspectives,'' {\em IEEE J. Sel. Topic Quantum. Electron.}~{\bf 16} (2010), 730-739

\bibitem{Kruger2010}
R.~A.~Kruger, R.~B.~Lam, D.~R.~Reinecke,  S.~P.~D.~Rio and  R.~P.~Doyle, ``Photoacoustic angiography of the breast,'' {\em Med. Phys.} ~{\bf 37}, (2010), 6096-6100 

\bibitem{Ermilov2009}
S.~A.~Ermilov, T.~Khamapirad,  A.~Conjusteau,  M.~H.~Leonard,  R.~Lacewell,  K.~Mehta, T.~Miller and  A.~A.~Oraevsky, ``Laser optoacoustic imaging system for detection of breast cancer,'' {\em J. Biomed. Opt.} ~{\bf 14}(2) (2009), 024007

\bibitem{Pramanik2008}
M.~Pramanik, G.~Ku, C.~Li and L.~V.~Wang, ``Design and evaluation of a novel breast cancer detection system combining both thermoacoustic (TA) and photoacoustic (PA) tomography,'' {\em Med. Phys.}~{\bf 35}, (2008), 2218-2223

\bibitem{Heijblom2012}
M. Heijblom, D. Piras, W. Xia, J.C.G. van Hespen, J.M. Klaase, F.M. van den Engh, T.G. van Leeuwen, W. Steenbergen, and S. Manohar, ``Visualizing breast cancer using the Twente photoacoustic mammoscope: What do we learn from twelve new patient measurements?,'' {\em Opt. Express} {\bf 20}, (2012), 11582-11597 (doi: http://dx.doi.org/10.1364/OE.20.011582)

\bibitem{Xi2012}
L. Xi, X. Li, L. Yao, S. Grobmyer, and H. Jiang, ``Design and evaluation of a hybrid photoacoustic tomography and diffuse optical tomography system for breast cancer detection,'' {\em Med. Phys.}, {\bf 39}, (2012), 2584-2594 

\bibitem{Xie2011}
Z. Xie, X. Wang, R. F. Morris, F. R. Padilla, G. L. Lecarpentier, and P. L. Carson, ``Photoacoustic imaging for deep targets in the breast using a multichannel 2D array transducer,'' {\em SPIE Proceedings},  {\bf 7899}, (2011), 1172-1181

\bibitem{Khokhlova2007}
T. D. Khokhlova, I. M. Pelivanov, V. V. Kozhushko, A. N. Zharinov, V. S. Solomatin and A. A. Karabutov, ``Optoacoustic imaging of absorbing objects in a turbid medium: ultimate sensitivity and application to breast cancer diagnostics,'' {\em Appl. Opt.}, {\bf 46}, (2007), 262-272 

\bibitem{Wang2007}
Y. Wang, T. N. Erpelding, L. Jankovic, Z. Guo, J. Robert, G. David and L. V. Wang, ``In vivo three-dimensional photoacoustic imaging based on a clinical matrix array ultrasound probe,'' {\em J. Biomed. Opt.}, {\bf 17}(6), (2012), 061208 

\bibitem{Xing2010}
F.~Ye, S.~Yang, and D.~Xing, ``Three-dimensional photoacoustic imaging system in line confocal mode for breast cancer detection,'' {\em Appl. Phys. Lett.}, {\bf 97}, (2010), 213702 doi:http://dx.doi.org/10.1063/1.3518704 

\bibitem{Pierce1989}
A.~D.~Pierce, ``Acoustics: An Introduction to its Physical Principles and Applications,''{\em Acoustic society of America, New York}, (1989) 

\bibitem{Xu2003}
M.~H.~Xu and L.~V.~Wang, ``Analytic explanation of spatial resolution related to bandwidth and detector aperture size in thermoacoustic or photoacoustic reconstruction,'' {\em Phys. Rev. E}~{\bf 67}, (2003), 056605

\bibitem{Li2008}
 C.~Li, G.~Ku and L.~V.~Wang, ``Negative lens concept for photoacoustic tomography,'' {\em Phys. Rev. E}~{\bf 78} (2008), 021901

\bibitem{Pramanik2009}
M.~Pramanik, G.~Ku and L.~V.~Wang, ``Tangential resolution improvement in thermoacoustic and photoacoustic tomography using a nagative acoustic lens,'' {\em J. Biomed. Opt.}~{\bf 14}(2), (2009), 024028

\bibitem{Brown2007}
J.~A.~Brown, F.~S.~Foster, A.~Needles, E.~Cherin and G.~R.~Lockwood, ``Fabrication and performance of a 40-MHz linear array based on a 1-3 composite with geometric elevation focusing,'' {\em IEEE Trans. Ultrason. Ferro Freq. Control}~{\bf 54}, (2007), 1888-1894

\bibitem{Xia2012}
W.~Xia, D.~Piras, J. C. G. van Hespen, S.~van~Veldhoven, C.~Prins, T.~G.~van~Leeuwen, W.~Steenbergen, and~S.~Manohar,``An optimized ultrasound detector for photoacoustic breast tomography,'' {\em Med. Phys.}~{\bf 40}(3) (2013), 032901

\bibitem{Bamber1997}
 J.~C.~Bamber, ``Acoustic characteristics of biological media,''{\em Encyclopedia of Acoustics} ~{\bf 4} ed M J Crocker (New York: Wiley) (chap 141), (1997), 1703--26
 
\bibitem{Xia2011}
W.~Xia, D.~Piras, M.~Heijblom, W.~Steenbergen, T.~G.~van Leeuwen and S.~Manohar, ``Poly(vinyl alcohol) gels as photoacoustic breast phantoms revisited,'' {\em J. Biomed. Opt.} ~{\bf 16}(7), (2011), 075002
 
\bibitem{Lubbers1998}
J.~Lubbers and R.~A.~Graaff, ``Simple and accurate formula for the sound velocity in water,'' {\em Ultrasound Med. Biol.}~{\bf 24}  (2008), 1065-1069

\bibitem{Sutherland1978}
H.~J.~Sutherland, ``Acoustical determination of the shear relaxation functions for polymethyl methacrylate and Epon 828-Z,'' {\em J. Appl. Phys.}~{\bf 48}, (1978), 3941-3945

\bibitem{Fish1990}
P.~Fish P, ``Physics and Instrumentation of Diagnostic Medical Ultrasound,'' (Chichester: Wiley), (1990)

\bibitem{Callerama1979}
J. Callerama, R. H. Tancrell and D. T. Wilson, ``Transmitters and receivers for medical ultrasonics'' Ultrasonic Symposium Proceedings, IEEE CH1482-9/79/0000-0407, (1979), 407-411

\bibitem{Greenspan1972}
M.~Greenspan, ``Acoustic properties of liquids,'' {\em American Institute of Physics (AIP) Handbook 3rd edition} ed D E Grey (New York: McGraw-Hill) p(3)-87, (1972)
 
\bibitem{Cox2005}
B.T. Cox and P.C. Beard, ``Fast calculation of pulsed photoacoustic field in fluids using k-space methods,'' {\em J. Acoust. Soc. Am.},  {\bf 117(6)}, (2005), 3616-3627 

\bibitem{Treeby2010s}
B.~E.~Treeby, and B.~Cox, ``k-Wave: MATLAB toolbox for the simulation and reconstruction of photoacoustic wave fields,'' {\em J. Biomed. Opt.}~{\bf 51}(2), (2010), 021314

\bibitem{Treeby2010}
B.~E.~Treeby, E.~Z.~Zhang and B.~Cox, ``Photoacoustic tomography in absorbing acoustic media using time reversal,'' {\em Inverse Problems}~{\bf 26},  (2010), 115003

\bibitem{Staveren1991}
H.~J.~Van Staveren, C.~J.~M.~Moes, J.~van Marie, S.~A.~Prahl and M.~J.~C.~van Gemert, ``Light scattering in Intralipid--10$\%$ in the wavelength range of 400-1100 nm,''{\em Appl. Opt.}~{\bf 30}, (1991), 4507-4514

\bibitem{Curcio1951}
J. A. Curcio and C. C. Petty, ``The near infrared absorption spectrum of liquid water'', {\em J. Acoust. Soc. Am.},  {\bf 41}(5), (1951), 302-304

\bibitem{Hale1973}
G. M. Hale and M. R. Querry, ``Optical constants of water in the 200-nm to 200-$\rm\mu$m wavelength region'', {\em Appl. Opt.},  {\bf 12}(3), (1973), 555-562

\bibitem{Martelli2007}
F. Martelli and G. Zaccanti ``Calibration of scattering and absorption properties of a liquid diffusive medium at NIR wavelengths. CW method,'' {\em Opt. Express} ~{\bf 15}(2), (2007), 486--500

\bibitem{Bloomfield2000}
 P.~E.~Bloomfield, W.~Lo and P.~A.~Lewin, ``Experimental study of the acoustical properties of polymers utilized to construct \textsc{PVDF} ultrasonic transducers and the acousto-electric properties of \textsc{PVDF} and \textsc{P(VDF/T}r\textsc{FE)} films,''  {\em IEEE Trans. Ultrason. Ferro Freq. Control}~{\bf 47}(6), (2000), 1397-1405
 
\bibitem{Carlson2003}
 B.~T.~Carlson, J.~van Deventer, A.~Scolan and C.~Carlander, ``Frequency and temperature dependence of acoustic properties of polymers used in pulse-echo systems,'' {\em Proceedings of IEEE Ultrasonics symposium, Honolulu, HI, USA}, (2003), 885-888
 
\bibitem{Waters2005}
K.~R.~Waters and J.~G.~Miller, ``Causality-imposed (Kramers-Kr$\rm \ddot{o}$nig) relationships between attenuation and dispersion,'' {\em IEEE Trans. Ultrason. Ferro Freq. Control}~{\bf 52}, (2005), 822-834

\bibitem{McKeighen1998}
R. E. McKeighen, ``Design guidelines for medical ultrasound arrays,'' {\em Proc. SPIE Int. Symp. Med. Imag.}, {\bf 3341},(1998), 2-4  (doi: http://dx.doi.org/10.1117/12.307992)

\end{thebibliography}
\end{document}